\documentclass[10pt,twocolumn,final,conference,letterpaper]{IEEEtran}
\usepackage{color}
\usepackage[latin1]{inputenc}
\usepackage{amsmath}
\usepackage{amsfonts}
\usepackage{units}
\usepackage{nicefrac}
\usepackage{subfigure}
\usepackage{cite}
\usepackage{multirow}	    
\usepackage{rotating}	    
\usepackage[printonlyused]{acronym}	    
\usepackage{pstricks, pst-plot, pst-grad, pstricks-add, pst-node, pstricks-add}

\newcommand{\Set}[1]{\mathcal{#1}}

\newcommand{\RV}[1]{\mathrm{#1}}

\newcommand{\C}[0]{\mathrm{C}}

\newcommand{\PathlossExponent}[0]{\alpha}

\newcommand{\CN}[0]{\mathcal{CN}}
\newcommand{\AvgThroughput}[0]{\ensuremath{\overline{\theta}}}
\newcommand{\FiveThroughput}[0]{\ensuremath{\theta_{5\%}}}

\acrodef{ARQ}[ARQ]{automatic repeat-request}
\acrodef{AWGN}[AWGN]{additive white Gaussian noise}
\acrodef{BS}[BS]{base station}
\acrodef{CDF}[CDF]{cumulative distribution function}
\acrodef{CSI}[CSI]{channel state information}
\acrodef{DL}[DL]{downlink}
\acrodef{DPC}[DPC]{dirty paper coding}
\acrodef{ETW}[ETW]{Etkin-Tse-Wang}
\acrodef{FDD}[FDD]{frequency division duplexing}
\acrodef{FDMA}[FDMA]{frequency division multiple access}
\acrodef{HK}[HK]{Han-Kobayashi}
\acrodef{IC}[IC]{Interference Channel}
\acrodef{LOS}[LOS]{line-of-sight}
\acrodef{MCS}[MCS]{modulation and coding schemes}
\acrodef{OFDM}[OFDM]{Orthogonal Frequency Division Multiplexing}
\acrodef{QoS}[QoS]{quality-of-service}
\acrodef{RAP}[RAP]{radio access point}
\acrodef{RN}[RN]{relay node}
\acrodef{SIC}[SIC]{successive interference cancelation}
\acrodef{SNR}[SNR]{signal-to-noise ratio}
\acrodef{SINR}[SINR]{signal-to-interference-and-noise ratio}
\acrodef{TDD}[TDD]{time division duplex}
\acrodef{TDMA}[TDMA]{time division multiple access}
\acrodef{UL}[UL]{uplink}
\acrodef{UT}[UT]{user terminal}
\acrodef{ZF-DPC}[ZF-DPC]{Zero-Forcing DPC}

\newrgbcolor{darkgreen}{0.2 0.6 0.2}
\newrgbcolor{darkred}{0.6 0.2 0.2}
\newrgbcolor{darkblue}{0.2 0.2 0.6}

\title{On the Transmission-Computation-Energy Tradeoff in Wireless and Fixed Networks}
\author{
\authorblockN{Peter Rost}%
\authorblockA{NEC Laboratories Europe\\Network  Divsion, Mobile and Wireless Networks Group\\69115 Heidelberg, Germany\\
Email: peter.rost@neclab.eu}
\and
\authorblockN{Gerhard Fettweis}
  \authorblockA{Technische Universität Dresden\\Vodafone Chair Mobile Communications Systems\\01069 Dresden, Germany\\
    Email: fettweis@ifn.et.tu-dresden.de
    }}%
\IEEEoverridecommandlockouts
\begin{document}
\maketitle
  \begin{abstract}
    In this paper, a framework for the analysis of the transmission-computation-energy tradeoff in wireless
    and fixed networks is introduced. 
    The analysis of this tradeoff considers both the transmission energy as well as the energy consumed at the receiver to process the
    received signal. While previous work considers linear decoder complexity, which is
    only achieved by uncoded transmission, this paper 
    claims that the average processing (or computation) energy per symbol depends exponentially on the 
    information rate of the source message. 
    The introduced framework is parametrized in a way that it reflects properties of fixed and wireless networks alike.
    
    The analysis of this paper shows that exponential complexity and therefore stronger codes are preferable at 
    low data rates while linear complexity and therefore uncoded transmission becomes preferable at high data rates.
    The more the computation energy is emphasized (such as in fixed networks), the less hops are optimal and the lower is the benefit of multi-hopping. 
    On the other hand, the higher the information rate of the single-hop network, the higher the benefits of multi-hopping.
    Both conclusions are underlined by analytical results.
  \end{abstract}
\IEEEpeerreviewmaketitle
%
%
%

  \section{Introduction}
    \subsection{Energy-Efficiency in Multi-Hop Networks}
      Recently WWRF Chair M.A. Uusitalo announced his vision of the future wireless world\footnote{Presentation is available
      on http://wireless-world-research.org/.}. One of his major technological visions is that until the year 2017 7 trillion
      wireless devices will be used by 7 billion users.
      Mobile communication engineers face a multitude of challenges to integrate this tremendous number of nodes
      such as more demanding requirements on the radio resource management, packet routing, and energy efficiency.
      The latter is in the focus of this paper, which analyzes the interplay of the energy consumption of the transceiver
      path (\emph{transmission energy}) and the data processing unit (\emph{computation energy}). We
      introduce a framework and draw conclusions, which can be equally applied to fixed networks, cellular networks, 
      and low-complexity sensor networks.

      There exists a comprehensive literature analyzing the transmission energy consumption in a wireless network such as the
      seminal work in \cite{Avestimehr.Tse.TransIT.2007}, where a bursty protocol has been introduced. A bursty protocol shortens the
      online time of a node, concentrates the transmitted energy on a shorter time interval, and increases the \ac{SNR} in analog
      multi-hop networks. Intuitively, a bursty transmission reduces the constant energy as the online-time of the node
      is reduced. On the other hand, the information rate is increased, which implies that the data processing unit
      potentially requires more energy, if the required average computation power per symbol scales super-linearly. 
      This inherent tradeoff is not only of importance for energy-limited terminals, which have only a limited energy budget, but
      it is of equal importance for fixed networks where data aggregation and bursty transmission are valid
      alternatives to reduce the energy consumption. Due to the tremendous number of sensor nodes and the requirement for
      a power source such as a battery, even small energy savings per node imply a significant `green potential'.
      This inherent tradeoff of transmission and computation energy is in the focus of this work and we explore 
      how an appropriate choice of the packet length and number of nodes in a network can reduce the overall energy consumption.

    \subsection{Related Work}
      In previous work such as \cite{Bhardwaj.Chandrakasan.ICC.2001,Melo.Liu.Globecom.2002} the computation energy
      of a network has been only considered under the assumption of a linear dependence of rate and energy.
      In \cite{Bhardwaj.Chandrakasan.ICC.2001}, the authors investigate the lifetime of a network where individual
      nodes collect and deliver data. In particular, it considers the transmission energy, the source behavior, network
      size, and also how much computation energy is required to receive a bit, which relates linearly to the information rate.
      Similarly, \cite{Melo.Liu.Globecom.2002} also analyzed the network lifetime and applied a linear model
      for the computation energy. As we discuss later, a linear model does not suitably reflect the case of coded
      transmission, since we rather face an exponential dependency. 
      The routing problem in wireless networks with per-bit processing-power has been analyzed in 
      \cite{Heinzelman.Chandrakasan.Balakrishnan.HICSS.2000}, where again the processing energy depends linearly on the information rate.

    \subsection{Contribution and Outline}
      This work introduces a framework to analyze and to assess the tradeoff of computation and transmission
      energy in multi-hop networks such as relay-based cellular networks, sensor networks, but also fixed networks with intermediate gateways and routers.
      We discuss the inter-play of both and show how packet length, data rate, network
      size, and the functional description of the computation complexity
      affects this relationship and depict potentials for the optimization of a network's energy consumption. The
      underlying system model will be introduced in Section \ref{sec:system.model}. Based on this model we derive
      the normalized computation and transmission energy of a decode-and-forward (also called store-and-forward)
      based multi-hop network in Section \ref{sec:protocol}.
      The tradeoff of both is illustrated in Section \ref{sec:results} and the paper is concluded in Section \ref{sec:conclusions}.

  \section{System Model}\label{sec:system.model}
    Before we present the transmission-computation-energy tradeoff in Section \ref{sec:protocol}, we successively introduce
    our channel model, resource model, and energy model.

    \subsection{Channel Model}
      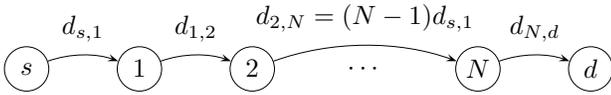
\begin{figure}
	\centering
	\begingroup
\unitlength=1mm
\begin{picture}(80, 14)(0, 0)
  \psset{xunit=1mm, yunit=1mm, linewidth=0.1mm}
  \psset{arrowsize=2pt 3, arrowlength=1.4, arrowinset=.4}
  \rput(0, 0){%
    \rput(0, 5){\cnode(0, 0){3mm}{S}\rput[c](0, 0){$s$}}%
    \rput(15, 5){\cnode(0, 0){3mm}{N1}\rput[c](0, 0){$1$}}\ncarc[arcangle=20, linestyle=solid]{->}{S}{N1}\naput{$d_{s, 1}$}%
    \rput(30, 5){\cnode(0, 0){3mm}{N2}\rput[c](0, 0){$2$}}\ncarc[arcangle=20, linestyle=solid]{->}{N1}{N2}\naput{$d_{1, 2}$}%
    \rput(45, 5){$\cdots$}%
    \rput(60, 5){\cnode(0, 0){3mm}{NN}\rput[c](0, 0){$N$}}\ncarc[arcangle=20, linestyle=solid]{->}{N2}{NN}\naput{$d_{2, N} = (N-1)d_{s, 1}$}%
    \rput(75, 5){\cnode(0, 0){3mm}{D}\rput[c](0, 0){$d$}}\ncarc[arcangle=20, linestyle=solid]{->}{NN}{D}\naput{$d_{N, d}$}%
  }
\end{picture}
\endgroup
	\caption{System setup with $N+2$ nodes distributed at equal distances.}
	\label{fig:model:system.setup}
      \end{figure}
      We consider a network composed of the source node $s=0$, the set of $N$ intermediate nodes $\Set{R}=[1; N]$, and the destination
      node $d=N+1$ as illustrated in Fig. \ref{fig:model:system.setup}. 
      This paper focuses on an \ac{AWGN} channel with fixed channel gain. The channel input
      at node $n$ is the complex Gaussian random process $\RV{X}_n\sim\CN(0, P_{tx, n})$ with average per-symbol power $P_{tx, n}$.
      Let the distance between two nodes $n, n'\in[0; N+1]$  be $d_{n, n'}$. Then the channel gain between both nodes 
      is $h_{n, n'} = d_{n, n'}^{-\PathlossExponent/2}$ with path loss exponent $\PathlossExponent$.
      For the sake of notational simplification we assume in this paper that all nodes are distributed at equal distance 
      between $s$ and $d$ such that 
      \begin{equation}
	d_{n, n'} = \frac{\left|n' - n\right|}{N+1}d_{s, d}.
      \end{equation}
      This assumption is rarely fulfilled in wireless sensor networks, but is of particular relevance in fixed networks. In addition, the
      conclusions drawn in this paper do not depend on this assumption but rather on the non-linear nature of path loss as well as the different
      complexity. In the \ac{AWGN} model, the channel output at node $n'$ is given by
      \begin{equation}
	\RV{Y}_{n'} = \sum\limits_{n\in[0; N+1]\setminus n'} h_{n', n}\RV{X}_n + \RV{Z}_{n'},
      \end{equation}
      where $\RV{Z}_{n'}\sim\CN(0, \sigma^2)$ is the \ac{AWGN} with $\sigma^2 = 1$ throughout this paper. 
      
      We consider in this work a network of full-duplex terminals in order to introduce the computation-transmission-energy tradeoff. Full-duplex
      is easily implemented in fixed networks where both links are physically separated using different physical cables. However, wireless
      applications imply a half-duplex constraint on the deployed nodes and therefore an inherent rate-loss, which renders
      multi-hop transmission less beneficial.

    \subsection{Resource Model and Means of Normalization}
      Our analysis compares the multi-hop setup with a single-hop reference system ($N=0$) with source-destination
      distance $d_\mathrm{ref}=d_{s,d}=1$, and the reference transmission power $P_{tx, s} = P_{tx, \mathrm{ref}}$. Let us assume that the source
      node has a fixed amount of data collected, which is mapped to the codeword $W_\mathrm{ref}$ 
      with overall $\left\|W_\mathrm{ref}\right\|$ bits net-data and
      must be delivered to the destination. Assume that $T_\mathrm{ref}<\infty$ exclusive resource
      elements (available channel uses) are assigned to the source node, which can correspond to time slots in an TDMA system or exclusive
      bandwidth in an FDMA system. Without loss of generality we refer in the following to $T_\mathrm{ref}$ as the number of symbols in time.
      In order to reliably communicate $W_\mathrm{ref}$, the source must transmit with an average information rate per symbol
      $R_\mathrm{ref} = \left\|W_\mathrm{ref}\right\|/T_\mathrm{ref}$, i.\,e.,
      with each channel access on average $R_\mathrm{ref}$ bits must be transmitted. In the previously described Gaussian \ac{AWGN} channel 
      model the average information rate per symbol is described by $R_\mathrm{ref} = \C(P_{tx, \mathrm{ref}}/\sigma^2)$ with $\C(x) = \log_2(1 + x)$. Throughout
      this paper, we use the number of bits $\left\|W_\mathrm{ref}\right\|$ as means of normalization and require that each protocol
      must reliably communicate $W_\mathrm{ref}$ using at most $T_\mathrm{ref}$ time symbols.

      This normalization offers the degree of freedom
      to adjust the number of used time symbols $T' < T_\mathrm{ref}$, which implies that a node uses only parts of the assigned
      resources. However, in order to deliver the same amount of data, the rate
      must be increased such that $R' = (T_\mathrm{ref}/T')R_\mathrm{ref}$ 
      as per channel access a higher number of information bits must be communicated.
      In the following part, we introduce our energy model and how the overall transmission and computation energy depend
      on the active time period $T'$.

    \subsection{Energy Model and Bursty Transmission}
      From the previous introduction we can immediately state that the transmission energy in the reference system is given by
      $E_{tx, \mathrm{ref}}=P_{tx, \mathrm{ref}}\cdot T_\mathrm{ref}$. Under the assumption that the 
      destination decodes the transmission also in
      a time interval $T_\mathrm{ref}$ (in order to avoid an accumulation of packets), it must also decode the data with rate $R_\mathrm{ref}$.
      Motivated by convolutional codes, which can be decoded using a trellis representation \cite{Tarokh.Seshadri.Calderbank.TransIT.1998}
      of the encoder's state space, we claim that the decoding complexity for each time symbol is exponential in the information
      rate $R_\mathrm{ref}$. This behavior is caused by
      the fact that also the state space and the number of possible state transitions per channel access in the 
      decoder-trellis expands exponentially with the product of constraint length and $R_\mathrm{ref}$. Previous work only considered
      linear complexity, which implies an uncoded transmission and an actual performance loss that can be expressed by
      a constant SNR gap as introduced in \cite[pp. 66]{Li.Stuber.2006}. 
      We apply an SNR gap between exponential and linear complexity of $\unit[5]{dB}$ \cite{Li.Stuber.2006, Yu.Varodayan.Cioffi.TransIT.2005}, which implies that
      a system with linear decoding complexity must invest $\unit[5]{dB}$ higher transmission power in order to achieve the same performance.

      The computation power can be expressed by $P_{c, \mathrm{ref}}\sim c_1c_3^{c_2R_\mathrm{ref}}$ where the constants $c_j$ are decoder specific.
      For the sake of simplicity, we consider in our work $c_1 = c_2 = 1$ (we neglect the constraint length as it remains the 
      same for all $T'$) and $c_3 = 2$ such that the computation energy for a packet
      of length $T_\mathrm{ref}$ is given by $E_{c, \mathrm{ref}} = T_\mathrm{ref}\cdot2^{R_\mathrm{ref}}$, which is used
      as reference value for our evaluation of multi-hop networks.
      The actual parametrization changes for different coding schemes and inherently affects the quantitative results
      of the transmission-computation-energy tradeoff although the qualitative conclusions are not affected.

      Assume the transmission length used by node $n$ is $T' < T_\mathrm{ref}$, then the transmission energy is scaled such that
      \begin{equation}
	T_\mathrm{ref}\C\left(\frac{P_{tx, n}}{\sigma^2}\right) = T'\C\left(\frac{P_{tx, n}'}{\sigma^2}\right)
      \end{equation}
      is fulfilled. This implies that the \emph{bursty} protocol with $T' < T_\mathrm{ref}$ requires the transmission power 
      \begin{equation}
	P_{tx, n}'(\delta_t = T' / T_\mathrm{ref}) = \sigma^2\left(\left(1+\frac{P_{tx, n}}{\sigma^2}\right)^{1/\delta_t} - 1\right)\label{eq:model:energy:tx.10}
      \end{equation}
      in order to satisfy the constraint that the same amount of data must be communicated. In addition, also the computation
      energy increases as the rate $R' = \C\left(\frac{P_{tx, n}'}{\sigma^2}\right)$ implies an exponential scaling of the
      computation energy. Let $\Delta_r=R'- R_\mathrm{ref}$, then the computation energy is given as
      \begin{equation}
	E_{c, n}' = 2^{\Delta_r} E_{c, n},\label{eq:model:energy:tx.11}
      \end{equation}
      which deviates from algorithms with linear complexity where $E_{c, n}'$ scales linearly with the ratio
      of $R'$ and $R_\mathrm{ref}$. In the following, all derivations are presented for exponential complexity while the
      corresponding equations for linear complexity can be easily obtained using a linear model in (\ref{eq:model:energy:tx.11}).

  \section{Normalized Energy in Multi-Hop Networks}\label{sec:protocol}
    In order to capture the tradeoff between transmission and computation energy in multi-hop networks, we use the decode-and-forward
    protocol introduced in \cite{Xie.Kumar.TransIT.2004} with node-cooperation and non-coherent transmission. Given the power assignment
    vector $[P_{tx, 0}, P_{tx, 1}, \dots P_{tx, N}]$ the maximum achievable end-to-end rate is given by \cite{Xie.Kumar.TransIT.2004}
    \begin{equation}
      R \leq \min\limits_{1\leq n\leq N+1} \C\left( \frac{1}{\sigma^2} \sum\limits_{k = 0}^{n - 1} h_{n, k}^2 P_{tx, k} \right).
    \end{equation}
    In the following, we define the normalized transmission and computation energy of a multi-hop network compared to a single-hop
    transmission for a fixed number of resource elements. On this basis, we extend the framework to define the normalized
    energy for a flexible and optimized number of resource elements in our multi-hop network.

    \subsection{Normalized Energy for Fixed $T_\mathrm{ref}$}
      The rate on the first hop is given by
      \begin{equation}
	R_0 = \C\left(\frac{P_{tx, 0}}{\sigma^2}\right)
      \end{equation}
      with the source power given as a function of the reference power:
      \begin{equation}
	P_{tx, 0} = (N+1)^{-\PathlossExponent}P_\mathrm{ref}.
      \end{equation}
      In order to achieve the same rate on the second hop, the transmission power of the second terminal must be chosen such that
      \begin{align}
	h_{0, 1}^2P_{tx, 0} & = h_{0, 1}^2P_{tx, 1} + h_{0, 2}^2P_{tx, 0}\text{ with }h_{n, n'} = d_{n, n'}^{-\PathlossExponent/2} \\
	P_{tx, 1} & = (1 - 2^{-\PathlossExponent})P_{tx, 0}.
      \end{align}
      This can be generalized for the transmission power of node $n$ as follows
      \begin{eqnarray}
       h_{0, 1}^2P_{tx, 0} & = & \sum\limits_{k = 0}^{n} h_{n+1, k}^2 P_{tx, k}\\ 
	& = & \sum\limits_{k = 0}^{n} (n+1-k)^{-\PathlossExponent}h_{0, 1}^2 P_{tx, k}\\
       P_{tx, n} & = & P_{tx, 0}  - \sum\limits_{k = 0}^{n - 1} (n+1-k)^{-\PathlossExponent} P_{tx, k}
      \end{eqnarray}
      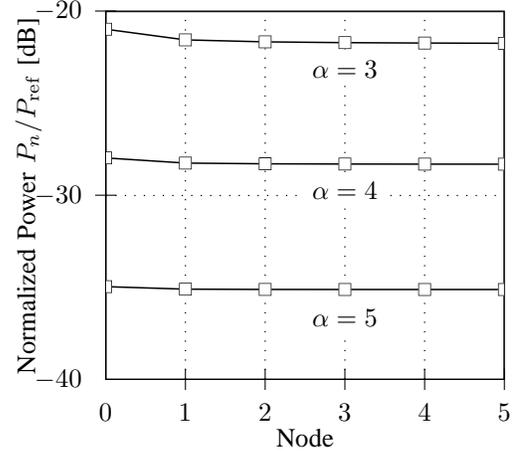
\begin{figure}
	\centering
	\begingroup
\unitlength=1mm
\psset{xunit=10.60000mm, yunit=2.45000mm, linewidth=0.1mm}
\psset{arrowsize=2pt 3, arrowlength=1.4, arrowinset=.4}\psset{axesstyle=frame}
\begin{pspicture}(-1.32075, -44.48980)(5.00000, -20.00000)
\rput(-0.18868, -0.81633){%
\psaxes[subticks=0, labels=all, xsubticks=1, ysubticks=1, Ox=0, Oy=-40, Dx=1, Dy=10]{-}(0.00000, -40.00000)(0.00000, -40.00000)(5.00000, -20.00000)%
\multips(1.00000, -40.00000)(1.00000, 0.0){4}{\psline[linecolor=black, linestyle=dotted, linewidth=0.2mm](0, 0)(0, 20.00000)}
\multips(0.00000, -30.00000)(0, 10.00000){1}{\psline[linecolor=black, linestyle=dotted, linewidth=0.2mm](0, 0)(5.00000, 0)}
\rput[b](2.50000, -43.67347){Node}
\rput[t]{90}(-1.13208, -30.00000){Normalized Power $P_n/P_\mathrm{ref}$ [dB]}
\psclip{\psframe(0.00000, -40.00000)(5.00000, -20.00000)}
\psline[linecolor=black, plotstyle=curve, linewidth=0.2mm, showpoints=true, linestyle=solid, dotstyle=square, dotscale=1.7 1.7](0.00000, -20.96910)(1.00000, -21.54902)(2.00000, -21.65662)(3.00000, -21.69915)(4.00000, -21.72086)(5.00000, -21.73358)
\psline[linecolor=black, plotstyle=curve, linewidth=0.2mm, showpoints=true, linestyle=solid, dotstyle=square, dotscale=1.7 1.7](0.00000, -27.95880)(1.00000, -28.23909)(2.00000, -28.27836)(3.00000, -28.29056)(4.00000, -28.29567)(5.00000, -28.29820)
\psline[linecolor=black, plotstyle=curve, linewidth=0.2mm, showpoints=true, linestyle=solid, dotstyle=square, dotscale=1.7 1.7](0.00000, -34.94850)(1.00000, -35.08638)(2.00000, -35.10048)(3.00000, -35.10385)(4.00000, -35.10499)(5.00000, -35.10546)
\endpsclip
\rput[t](3.00000, -22.20000){\psframebox[linestyle=none, fillcolor=white, fillstyle=solid]{$\PathlossExponent=3$}}
\rput[t](3.00000, -28.80000){\psframebox[linestyle=none, fillcolor=white, fillstyle=solid]{$\PathlossExponent=4$}}
\rput[t](3.00000, -35.70000){\psframebox[linestyle=none, fillcolor=white, fillstyle=solid]{$\PathlossExponent=5$}}
}\end{pspicture}
\endgroup
 
	\caption{Power assignment example for $5$ nodes and different path loss values}
	\label{fig:protocol:power.assignment.example}
      \end{figure}
      $P_{tx, n}$ is strictly monotonically decreasing in $n$ as illustrated in Fig. \ref{fig:protocol:power.assignment.example}.
      Hence, the power assignment $P_{tx, n} = P_{tx, 0}$ provides an achievable but suboptimal solution for wireless networks
      (in case of $\PathlossExponent=4$ the maximum difference in Fig. \ref{fig:protocol:power.assignment.example}
      is about $\unit[0.35]{dB}$ per node) and provides the exact solution for fixed networks where no cooperation gain can be exploited.
      The normalized network-wide transmission energy for packet length $T_\mathrm{ref}$ is
      \begin{align}
	E_\mathrm{tx, norm}(\delta_t = 1) & = \frac{T_\mathrm{ref}\sum\limits_{k=0}^N P_{tx, k}}{T_\mathrm{ref}P_\mathrm{ref}}
	  \leq \frac{(N+1)P_{tx, 0}}{P_\mathrm{ref}}\\
	  & = (N+1)^{1-\PathlossExponent}
      \end{align}
      and the normalized computation energy is given by
      \begin{eqnarray}
	E_\mathrm{c, norm}(\delta_t = 1) & = & \frac{T_\mathrm{ref}\sum\limits_{n=1}^{N+1} 2^{R_\mathrm{ref}}}{T_\mathrm{ref}2^{R_\mathrm{ref}}} = N + 1,
      \end{eqnarray}
      which already shows that the computation energy grows faster in $N$ than the transmission energy and therefore eventually
      becomes the dominant term for large $N$.

    \subsection{Bursty Transmission using $T' < T_\mathrm{ref}$}
      \begin{figure}
	\centering
	\begingroup
\unitlength=1mm
\psset{xunit=56.00000mm, yunit=4.90000mm, linewidth=0.1mm}
\psset{arrowsize=2pt 3, arrowlength=1.4, arrowinset=.4}\psset{axesstyle=frame}
\begin{pspicture}(-0.19643, -2.24490)(1.00000, 10.00000)
\rput(-0.03571, -0.40816){%
\psaxes[subticks=0, labels=all, xsubticks=1, ysubticks=1, Ox=0, Oy=0, Dx=0.2, Dy=2]{-}(0.00000, 0.00000)(0.00000, 0.00000)(1.00000, 10.00000)%
\multips(0.20000, 0.00000)(0.20000, 0.0){4}{\psline[linecolor=black, linestyle=dotted, linewidth=0.2mm](0, 0)(0, 10.00000)}
\multips(0.00000, 2.00000)(0, 2.00000){4}{\psline[linecolor=black, linestyle=dotted, linewidth=0.2mm](0, 0)(1.00000, 0)}
\rput[b](0.50000, -1.83673){$\delta_t$}
\rput[t]{90}(-0.16071, 5.00000){$E_\mathrm{tx, norm}$}
\psclip{\psframe(0.00000, 0.00000)(1.00000, 10.00000)}
\psline[linecolor=darkblue, plotstyle=curve, linewidth=0.2mm, showpoints=false, linestyle=solid, dotstyle=square, dotscale=1.7 1.7](0.00010, 11.00000)(0.01010, 11.00000)(0.02010, 11.00000)(0.03010, 11.00000)(0.04010, 11.00000)(0.05010, 11.00000)(0.06010, 11.00000)(0.07010, 11.00000)(0.08010, 11.00000)(0.09010, 11.00000)(0.10010, 11.00000)(0.11010, 11.00000)(0.12010, 11.00000)(0.13010, 11.00000)(0.14010, 11.00000)(0.15010, 11.00000)(0.16010, 11.00000)(0.17010, 9.83976)(0.18010, 8.27218)(0.19010, 7.09637)(0.20010, 6.19202)(0.21010, 5.48131)(0.22010, 4.91225)(0.23010, 4.44912)(0.24010, 4.06677)(0.25010, 3.74707)(0.26010, 3.47673)(0.27010, 3.24581)(0.28010, 3.04677)(0.29010, 2.87381)(0.30010, 2.72239)(0.31010, 2.58893)(0.32010, 2.47058)(0.33010, 2.36504)(0.34010, 2.27043)(0.35010, 2.18522)(0.36010, 2.10813)(0.37010, 2.03811)(0.38010, 1.97426)(0.39010, 1.91585)(0.40010, 1.86223)(0.41010, 1.81285)(0.42010, 1.76726)(0.43010, 1.72504)(0.44010, 1.68586)(0.45010, 1.64940)(0.46010, 1.61540)(0.47010, 1.58363)(0.48010, 1.55388)(0.49010, 1.52597)(0.50010, 1.49975)(0.51010, 1.47506)(0.52010, 1.45178)(0.53010, 1.42979)(0.54010, 1.40900)(0.55010, 1.38931)(0.56010, 1.37064)(0.57010, 1.35291)(0.58010, 1.33605)(0.59010, 1.32001)(0.60010, 1.30473)(0.61010, 1.29016)(0.62010, 1.27624)(0.63010, 1.26294)(0.64010, 1.25021)(0.65010, 1.23803)(0.66010, 1.22636)(0.67010, 1.21516)(0.68010, 1.20441)(0.69010, 1.19409)(0.70010, 1.18416)(0.71010, 1.17462)(0.72010, 1.16542)(0.73010, 1.15657)(0.74010, 1.14803)(0.75010, 1.13980)(0.76010, 1.13185)(0.77010, 1.12418)(0.78010, 1.11677)(0.79010, 1.10960)(0.80010, 1.10266)(0.81010, 1.09595)(0.82010, 1.08945)(0.83010, 1.08316)(0.84010, 1.07706)(0.85010, 1.07114)(0.86010, 1.06540)(0.87010, 1.05983)(0.88010, 1.05442)(0.89010, 1.04916)(0.90010, 1.04406)(0.91010, 1.03909)(0.92010, 1.03427)(0.93010, 1.02957)(0.94010, 1.02500)(0.95010, 1.02055)(0.96010, 1.01622)(0.97010, 1.01200)(0.98010, 1.00788)(0.99010, 1.00387)
\psline[linecolor=darkblue, plotstyle=curve, linewidth=0.2mm, showpoints=true, linestyle=none, dotstyle=square, dotscale=1.7 1.7](0.00010, 11.00000)(0.05010, 11.00000)(0.10010, 11.00000)(0.15010, 11.00000)(0.20010, 6.19202)(0.25010, 3.74707)(0.30010, 2.72239)(0.35010, 2.18522)(0.40010, 1.86223)(0.45010, 1.64940)(0.50010, 1.49975)(0.55010, 1.38931)(0.60010, 1.30473)(0.65010, 1.23803)(0.70010, 1.18416)(0.75010, 1.13980)(0.80010, 1.10266)(0.85010, 1.07114)(0.90010, 1.04406)(0.95010, 1.02055)
\psline[linecolor=darkred, plotstyle=curve, linewidth=0.2mm, showpoints=false, linestyle=solid, dotstyle=square, dotscale=1.7 1.7](0.00010, 11.00000)(0.01010, 11.00000)(0.02010, 11.00000)(0.03010, 2.39300)(0.04010, 1.20992)(0.05010, 0.82338)(0.06010, 0.64439)(0.07010, 0.54417)(0.08010, 0.48102)(0.09010, 0.43794)(0.10010, 0.40681)(0.11010, 0.38334)(0.12010, 0.36505)(0.13010, 0.35042)(0.14010, 0.33845)(0.15010, 0.32849)(0.16010, 0.32008)(0.17010, 0.31288)(0.18010, 0.30665)(0.19010, 0.30122)(0.20010, 0.29643)(0.21010, 0.29218)(0.22010, 0.28838)(0.23010, 0.28497)(0.24010, 0.28189)(0.25010, 0.27909)(0.26010, 0.27654)(0.27010, 0.27420)(0.28010, 0.27206)(0.29010, 0.27008)(0.30010, 0.26825)(0.31010, 0.26655)(0.32010, 0.26498)(0.33010, 0.26350)(0.34010, 0.26213)(0.35010, 0.26084)(0.36010, 0.25963)(0.37010, 0.25850)(0.38010, 0.25742)(0.39010, 0.25641)(0.40010, 0.25546)(0.41010, 0.25455)(0.42010, 0.25370)(0.43010, 0.25288)(0.44010, 0.25211)(0.45010, 0.25137)(0.46010, 0.25067)(0.47010, 0.25000)(0.48010, 0.24936)(0.49010, 0.24875)(0.50010, 0.24816)(0.51010, 0.24760)(0.52010, 0.24707)(0.53010, 0.24655)(0.54010, 0.24605)(0.55010, 0.24558)(0.56010, 0.24512)(0.57010, 0.24468)(0.58010, 0.24425)(0.59010, 0.24385)(0.60010, 0.24345)(0.61010, 0.24307)(0.62010, 0.24270)(0.63010, 0.24235)(0.64010, 0.24200)(0.65010, 0.24167)(0.66010, 0.24135)(0.67010, 0.24104)(0.68010, 0.24073)(0.69010, 0.24044)(0.70010, 0.24016)(0.71010, 0.23988)(0.72010, 0.23961)(0.73010, 0.23935)(0.74010, 0.23910)(0.75010, 0.23886)(0.76010, 0.23862)(0.77010, 0.23838)(0.78010, 0.23816)(0.79010, 0.23794)(0.80010, 0.23772)(0.81010, 0.23752)(0.82010, 0.23731)(0.83010, 0.23711)(0.84010, 0.23692)(0.85010, 0.23673)(0.86010, 0.23655)(0.87010, 0.23637)(0.88010, 0.23619)(0.89010, 0.23602)(0.90010, 0.23586)(0.91010, 0.23569)(0.92010, 0.23553)(0.93010, 0.23538)(0.94010, 0.23522)(0.95010, 0.23507)(0.96010, 0.23493)(0.97010, 0.23478)(0.98010, 0.23464)(0.99010, 0.23451)
\psline[linecolor=darkred, plotstyle=curve, linewidth=0.2mm, showpoints=true, linestyle=none, dotstyle=square, dotscale=1.7 1.7](0.00010, 11.00000)(0.05010, 0.95146)(0.10010, 0.44915)(0.15010, 0.35776)(0.20010, 0.32076)(0.25010, 0.30087)(0.30010, 0.28848)(0.35010, 0.28004)(0.40010, 0.27391)(0.45010, 0.26926)(0.50010, 0.26562)(0.55010, 0.26269)(0.60010, 0.26027)(0.65010, 0.25825)(0.70010, 0.25654)(0.75010, 0.25507)(0.80010, 0.25379)(0.85010, 0.25267)(0.90010, 0.25167)(0.95010, 0.25079)
\endpsclip
\psframe[linecolor=black, fillstyle=solid, fillcolor=white, shadowcolor=lightgray, shadowsize=1mm, shadow=true](0.44643, 6.12245)(0.83929, 8.16327)
\rput[l](0.60714, 7.55102){$N = 0$}
\psline[linecolor=darkblue, linestyle=solid, linewidth=0.3mm](0.48214, 7.55102)(0.55357, 7.55102)
\psline[linecolor=darkblue, linestyle=solid, linewidth=0.3mm](0.48214, 7.55102)(0.55357, 7.55102)
\psdots[linecolor=darkblue, linestyle=solid, linewidth=0.3mm, dotstyle=square, dotscale=1.7 1.7, linecolor=darkblue](0.51786, 7.55102)
\rput[l](0.60714, 6.73469){$N = 1$}
\psline[linecolor=darkred, linestyle=solid, linewidth=0.3mm](0.48214, 6.73469)(0.55357, 6.73469)
\psline[linecolor=darkred, linestyle=solid, linewidth=0.3mm](0.48214, 6.73469)(0.55357, 6.73469)
\psdots[linecolor=darkred, linestyle=solid, linewidth=0.3mm, dotstyle=square, dotscale=1.7 1.7, linecolor=darkred](0.51786, 6.73469)
}\end{pspicture}
\endgroup
 
	\caption{Normalized transmission energy for different relative $\delta_t$ and number of relays $N$.
	Markers indicate the approximation derived in (\ref{eq:protocol:400}) and lines give the exact solution
	as derived in (\ref{eq:protocol:390}). This example uses $\PathlossExponent=3$, $P_{ref} = 1$, $\sigma^2=1$,
	which implies $R_\mathrm{ref} = 1$.}
	\label{fig:normalized.transmission.energy.example}
      \end{figure}
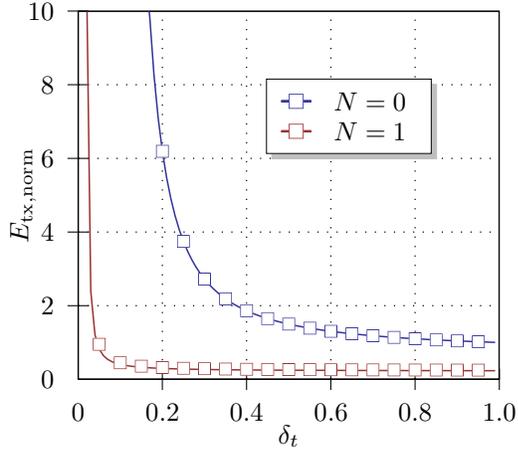
      If the packet length used by the multi-hop protocol is chosen as $T' < T_\mathrm{ref}$ for each hop, 
      then the normalized transmission energy is according to (\ref{eq:model:energy:tx.10}) and with $\delta_t = T' / T_\mathrm{ref}$
      given by
      \begin{align}
	E_\mathrm{tx, norm}(\delta_t) & = \delta_t\frac{\sum\limits_{n=0}^N P_{tx, n}'(\delta_t)}{P_\mathrm{tx, ref}}\\
	  & = \delta_t\frac{\sum\limits_{n=0}^N\sigma^2\left(\left(1+\frac{P_{tx, n}}{\sigma^2}\right)^{1/\delta_t} - 1\right)}{P_\mathrm{tx, ref}}\label{eq:protocol:390}\\
	  & \leq \delta_t(N+1)^{1-\PathlossExponent}
	  \frac{\sigma^2\left(\left(1+\frac{P_{tx, 0}}{\sigma^2}\right)^{1/\delta_t} - 1\right)}{P_{tx, 0}}.
	  \label{eq:protocol:400}
      \end{align}
      An example for $E_\mathrm{tx, norm}(\delta_t)$
      is shown in Fig. \ref{fig:normalized.transmission.energy.example} for $\PathlossExponent=3$, $P_{tx, \mathrm{ref}} = 1$,
      and $\sigma^2=1$, which gives $R_\mathrm{ref} = 1$.

      \begin{figure}
	\centering
	\begingroup
\unitlength=1mm
\psset{xunit=56.00000mm, yunit=4.90000mm, linewidth=0.1mm}
\psset{arrowsize=2pt 3, arrowlength=1.4, arrowinset=.4}\psset{axesstyle=frame}
\begin{pspicture}(-0.19643, -2.24490)(1.00000, 10.00000)
\rput(-0.03571, -0.40816){%
\psaxes[subticks=0, labels=all, xsubticks=1, ysubticks=1, Ox=0, Oy=0, Dx=0.2, Dy=2]{-}(0.00000, 0.00000)(0.00000, 0.00000)(1.00000, 10.00000)%
\multips(0.20000, 0.00000)(0.20000, 0.0){4}{\psline[linecolor=black, linestyle=dotted, linewidth=0.2mm](0, 0)(0, 10.00000)}
\multips(0.00000, 2.00000)(0, 2.00000){4}{\psline[linecolor=black, linestyle=dotted, linewidth=0.2mm](0, 0)(1.00000, 0)}
\rput[b](0.50000, -1.83673){$\delta_t$}
\rput[t]{90}(-0.16071, 5.00000){$E_\mathrm{c, norm}$}
\psclip{\psframe(0.00000, 0.00000)(1.00000, 10.00000)}
\psline[linecolor=darkblue, plotstyle=curve, linewidth=0.2mm, showpoints=false, linestyle=solid, dotstyle=square, dotscale=1.7 1.7](0.00010, 11.00000)(0.01010, 11.00000)(0.02010, 11.00000)(0.03010, 11.00000)(0.04010, 11.00000)(0.05010, 11.00000)(0.06010, 11.00000)(0.07010, 11.00000)(0.08010, 11.00000)(0.09010, 11.00000)(0.10010, 11.00000)(0.11010, 11.00000)(0.12010, 11.00000)(0.13010, 11.00000)(0.14010, 9.86478)(0.15010, 7.60117)(0.16010, 6.07608)(0.17010, 5.00493)(0.18010, 4.22614)(0.19010, 3.64323)(0.20010, 3.19606)(0.21010, 2.84571)(0.22010, 2.56618)(0.23010, 2.33961)(0.24010, 2.15343)(0.25010, 1.99858)(0.26010, 1.86841)(0.27010, 1.75795)(0.28010, 1.66344)(0.29010, 1.58196)(0.30010, 1.51125)(0.31010, 1.44952)(0.32010, 1.39534)(0.33010, 1.34757)(0.34010, 1.30526)(0.35010, 1.26766)(0.36010, 1.23411)(0.37010, 1.20410)(0.38010, 1.17718)(0.39010, 1.15297)(0.40010, 1.13116)(0.41010, 1.11148)(0.42010, 1.09368)(0.43010, 1.07757)(0.44010, 1.06298)(0.45010, 1.04975)(0.46010, 1.03775)(0.47010, 1.02687)(0.48010, 1.01699)(0.49010, 1.00804)(0.50010, 0.99992)(0.51010, 0.99258)(0.52010, 0.98594)(0.53010, 0.97995)(0.54010, 0.97455)(0.55010, 0.96970)(0.56010, 0.96537)(0.57010, 0.96150)(0.58010, 0.95808)(0.59010, 0.95506)(0.60010, 0.95242)(0.61010, 0.95013)(0.62010, 0.94817)(0.63010, 0.94652)(0.64010, 0.94516)(0.65010, 0.94407)(0.66010, 0.94323)(0.67010, 0.94263)(0.68010, 0.94226)(0.69010, 0.94209)(0.70010, 0.94213)(0.71010, 0.94236)(0.72010, 0.94276)(0.73010, 0.94333)(0.74010, 0.94407)(0.75010, 0.94495)(0.76010, 0.94598)(0.77010, 0.94714)(0.78010, 0.94843)(0.79010, 0.94985)(0.80010, 0.95138)(0.81010, 0.95303)(0.82010, 0.95478)(0.83010, 0.95663)(0.84010, 0.95858)(0.85010, 0.96062)(0.86010, 0.96275)(0.87010, 0.96496)(0.88010, 0.96726)(0.89010, 0.96963)(0.90010, 0.97208)(0.91010, 0.97460)(0.92010, 0.97718)(0.93010, 0.97984)(0.94010, 0.98255)(0.95010, 0.98533)(0.96010, 0.98816)(0.97010, 0.99105)(0.98010, 0.99399)(0.99010, 0.99699)
\psline[linecolor=darkblue, plotstyle=curve, linewidth=0.2mm, showpoints=true, linestyle=none, dotstyle=square, dotscale=1.7 1.7](0.00010, 11.00000)(0.05010, 11.00000)(0.10010, 11.00000)(0.15010, 7.60117)(0.20010, 3.19606)(0.25010, 1.99858)(0.30010, 1.51125)(0.35010, 1.26766)(0.40010, 1.13116)(0.45010, 1.04975)(0.50010, 0.99992)(0.55010, 0.96970)(0.60010, 0.95242)(0.65010, 0.94407)(0.70010, 0.94213)(0.75010, 0.94495)(0.80010, 0.95138)(0.85010, 0.96062)(0.90010, 0.97208)(0.95010, 0.98533)
\psline[linecolor=darkred, plotstyle=curve, linewidth=0.2mm, showpoints=false, linestyle=solid, dotstyle=square, dotscale=1.7 1.7](0.00010, 11.00000)(0.01010, 11.00000)(0.02010, 11.00000)(0.03010, 11.00000)(0.04010, 11.00000)(0.05010, 11.00000)(0.06010, 11.00000)(0.07010, 11.00000)(0.08010, 11.00000)(0.09010, 11.00000)(0.10010, 11.00000)(0.11010, 11.00000)(0.12010, 11.00000)(0.13010, 11.00000)(0.14010, 11.00000)(0.15010, 11.00000)(0.16010, 11.00000)(0.17010, 11.00000)(0.18010, 8.45228)(0.19010, 7.28647)(0.20010, 6.39212)(0.21010, 5.69141)(0.22010, 5.13235)(0.23010, 4.67922)(0.24010, 4.30687)(0.25010, 3.99717)(0.26010, 3.73683)(0.27010, 3.51591)(0.28010, 3.32687)(0.29010, 3.16391)(0.30010, 3.02249)(0.31010, 2.89903)(0.32010, 2.79068)(0.33010, 2.69514)(0.34010, 2.61053)(0.35010, 2.53532)(0.36010, 2.46823)(0.37010, 2.40821)(0.38010, 2.35436)(0.39010, 2.30595)(0.40010, 2.26233)(0.41010, 2.22295)(0.42010, 2.18736)(0.43010, 2.15514)(0.44010, 2.12596)(0.45010, 2.09950)(0.46010, 2.07550)(0.47010, 2.05373)(0.48010, 2.03398)(0.49010, 2.01607)(0.50010, 1.99985)(0.51010, 1.98516)(0.52010, 1.97188)(0.53010, 1.95989)(0.54010, 1.94910)(0.55010, 1.93941)(0.56010, 1.93074)(0.57010, 1.92301)(0.58010, 1.91615)(0.59010, 1.91011)(0.60010, 1.90483)(0.61010, 1.90026)(0.62010, 1.89634)(0.63010, 1.89304)(0.64010, 1.89031)(0.65010, 1.88813)(0.66010, 1.88646)(0.67010, 1.88526)(0.68010, 1.88451)(0.69010, 1.88419)(0.70010, 1.88426)(0.71010, 1.88472)(0.72010, 1.88552)(0.73010, 1.88667)(0.74010, 1.88813)(0.75010, 1.88990)(0.76010, 1.89195)(0.77010, 1.89428)(0.78010, 1.89687)(0.79010, 1.89970)(0.80010, 1.90276)(0.81010, 1.90605)(0.82010, 1.90955)(0.83010, 1.91326)(0.84010, 1.91716)(0.85010, 1.92124)(0.86010, 1.92550)(0.87010, 1.92993)(0.88010, 1.93452)(0.89010, 1.93926)(0.90010, 1.94416)(0.91010, 1.94919)(0.92010, 1.95437)(0.93010, 1.95967)(0.94010, 1.96510)(0.95010, 1.97065)(0.96010, 1.97632)(0.97010, 1.98210)(0.98010, 1.98798)(0.99010, 1.99397)
\psline[linecolor=darkred, plotstyle=curve, linewidth=0.2mm, showpoints=true, linestyle=none, dotstyle=square, dotscale=1.7 1.7](0.00010, 11.00000)(0.05010, 11.00000)(0.10010, 11.00000)(0.15010, 11.00000)(0.20010, 6.39212)(0.25010, 3.99717)(0.30010, 3.02249)(0.35010, 2.53532)(0.40010, 2.26233)(0.45010, 2.09950)(0.50010, 1.99985)(0.55010, 1.93941)(0.60010, 1.90483)(0.65010, 1.88813)(0.70010, 1.88426)(0.75010, 1.88990)(0.80010, 1.90276)(0.85010, 1.92124)(0.90010, 1.94416)(0.95010, 1.97065)
\endpsclip
\psframe[linecolor=black, fillstyle=solid, fillcolor=white, shadowcolor=lightgray, shadowsize=1mm, shadow=true](0.44643, 6.12245)(0.83929, 8.16327)
\rput[l](0.60714, 7.55102){$N = 0$}
\psline[linecolor=darkblue, linestyle=solid, linewidth=0.3mm](0.48214, 7.55102)(0.55357, 7.55102)
\psline[linecolor=darkblue, linestyle=solid, linewidth=0.3mm](0.48214, 7.55102)(0.55357, 7.55102)
\psdots[linecolor=darkblue, linestyle=solid, linewidth=0.3mm, dotstyle=square, dotscale=1.7 1.7, linecolor=darkblue](0.51786, 7.55102)
\rput[l](0.60714, 6.73469){$N = 1$}
\psline[linecolor=darkred, linestyle=solid, linewidth=0.3mm](0.48214, 6.73469)(0.55357, 6.73469)
\psline[linecolor=darkred, linestyle=solid, linewidth=0.3mm](0.48214, 6.73469)(0.55357, 6.73469)
\psdots[linecolor=darkred, linestyle=solid, linewidth=0.3mm, dotstyle=square, dotscale=1.7 1.7, linecolor=darkred](0.51786, 6.73469)
}\end{pspicture}
\endgroup
 
	\caption{Normalized transmission energy for different relative packet lengths using the expression in (\ref{eq:protocol:420})
	and for reference rate $R_\mathrm{ref} = 1$.}
	\label{fig:normalized.computation.energy.example}
      \end{figure}
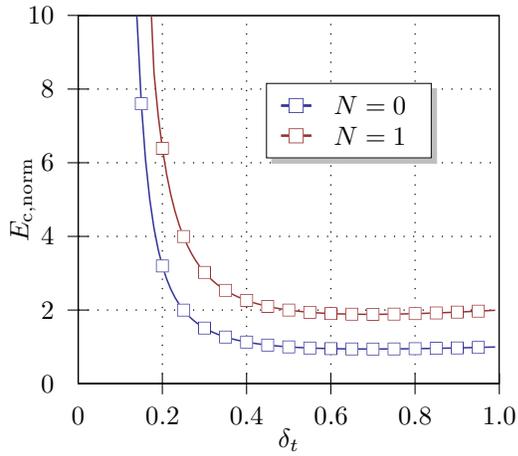
      Similarly, the computation energy is given by
      \begin{align}
	E_\mathrm{c, norm}(\delta_t) & = \delta_t \frac{\sum\limits_{n=1}^{N+1}2^{R_\mathrm{ref}}2^{\Delta_r}}{2^{R_\mathrm{ref}}} \\
	  & = \delta_t(N+1)2^{\Delta_r} = \delta_t2^{\Delta_r}E_\mathrm{c, norm}(1).
	  \label{eq:protocol:410}
      \end{align}
      The rate difference $\Delta_r = R' - R_\mathrm{ref}$ must be such that $R_\mathrm{ref}T_\mathrm{ref} = R'T'$, which implies
      \begin{align}
	E_\mathrm{c, norm}(\delta_t) & = \delta_t 2^{R_\mathrm{ref}(1/\delta_t - 1)}E_\mathrm{c, norm}(1).
	  \label{eq:protocol:420}
      \end{align}
      An example for the normalized computation energy is shown in Fig. \ref{fig:normalized.computation.energy.example} 
      for the reference rate $R_\mathrm{ref} = 1$.
      Interestingly, assume $T_\mathrm{ref} = 2T'$ ($\delta_t = 0.5$) 
      and $R_\mathrm{ref} < 1$ then a bursty transmission will not consume more computation energy
      than the direct transmission. On the other hand, if the rate increases to $R_\mathrm{ref} > 1$, the bursty transmission will
      increase the required computation energy.  

    \subsection{Optimal Packet Length $T'$}\label{sec:protocol:optimum}  
      We can easily identify the packet length, which minimizes the computation energy in the multi-hop network
      (once we chose $N$) to be
      \begin{align}
	T_\mathrm{c, opt}' & = \arg\min\limits_{0<T'\leq T} E_\mathrm{c, norm}(T'/T_\mathrm{ref}) \\
	\frac{\mathrm{d}E_\mathrm{c, norm}(T'/T_\mathrm{ref})}{\mathrm{d}T'} & \stackrel{!}{=} 0 \\
	T_\mathrm{c, opt}' & = T_\mathrm{ref}\min\left((\ln 2) R_\mathrm{ref}, 1\right).\label{eq:protocol:520}
      \end{align}

      We know from (\ref{eq:protocol:400}) that the transmission energy can only increase for $T' < T_\mathrm{ref}$ such that the optimal
      $T'$, which minimizes the overall energy consumption, must be in the interval $[T_\mathrm{c, opt}'; T_\mathrm{ref}]$ and depends on
      the ratio of $E_\mathrm{tx, norm}$ and $E_\mathrm{c, norm}$. However, $T_\mathrm{c, opt}'$ provides a good lower bound on
      the optimal packet length and only depends on the reference rate $R_\mathrm{ref}$, which simplifies its computation.

      The overall energy required by the network is given as
      \begin{align}
	E_\mathrm{sum}(R_\mathrm{ref}) & = E_\mathrm{c, norm}(R_\mathrm{ref}) \cdot E_\mathrm{c, ref}(R_\mathrm{ref})\nonumber\\
	  &\quad{+}\; E_\mathrm{tx, norm}(R_\mathrm{ref})\cdot E_\mathrm{tx, ref}(R_\mathrm{ref}).
      \end{align}
      Now let $E_\mathrm{c, ref} = \eta_\mathrm{ref}(R_\mathrm{ref})E_\mathrm{tx, ref}$,
      which relates the computation and transmission energy
      for a single-hop system depending on the actual rate. Using (\ref{eq:model:energy:tx.10}) and (\ref{eq:model:energy:tx.11}),
      the function $\eta_\mathrm{ref}(R_\mathrm{ref})$ can be expressed depending on a system-specific reference
      value $\eta_\mathrm{ref}(1)$ (as shown in the appendix).
      The overall consumed energy in the multi-hop network is now given as
      \begin{align}
	E_\mathrm{sum}(R_\mathrm{ref}) & = E_\mathrm{c, norm} \cdot \eta_\mathrm{ref}(R_\mathrm{ref})E_\mathrm{tx, ref}\nonumber\\
	  &\quad{+}\; E_\mathrm{tx, norm}\cdot E_\mathrm{tx, ref},
      \end{align}
      where $E_\mathrm{c, norm}$ and $E_\mathrm{tx, ref}$ are the normalized energies for reference rate $R_\mathrm{ref}$ (which
      are omitted here to avoid any confusion with $\delta_t$). The normalized sum-energy can be expressed by
      \begin{align}
	E_\mathrm{sum, norm}(R_\mathrm{ref}) & = \frac{E_\mathrm{c, norm} \cdot \eta_\mathrm{ref}(R_\mathrm{ref}) + E_\mathrm{tx, norm}}{1 + \eta_\mathrm{ref}(R_\mathrm{ref}))}.\label{eq:protocol:590}
      \end{align}
    
  \section{Results}\label{sec:results}
    Based on the previously described framework, we discuss in this section results for the transmission-computation-energy tradeoff. 
    We focus thereby on two particular aspects. Firstly, we analyze the jointly achievable transmission-computation-energy curve under
    a given throughput-constraint, and secondly, we discuss the optimal network size and energy savings potential 
    depending on the reference rate $R_\mathrm{ref}$.

-- Strange point at 1.5 value of Rref
    
    \subsection{Transmission-Computation-Energy Tradeoff}      
      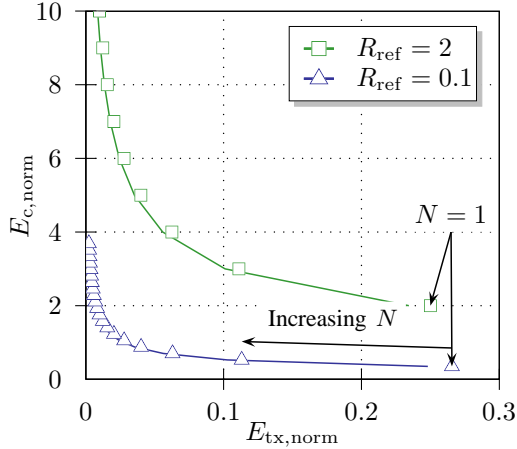
\begin{figure}
	\centering
	\begingroup
\unitlength=1mm
\psset{xunit=183.27224mm, yunit=4.90000mm, linewidth=0.1mm}
\psset{arrowsize=2pt 3, arrowlength=1.4, arrowinset=.4}\psset{axesstyle=frame}
\begin{pspicture}(-0.06548, -2.24490)(0.30010, 10.00000)
\rput(-0.01091, -0.40816){%
\psaxes[subticks=0, labels=all, xsubticks=1, ysubticks=1, Ox=0, Oy=0, Dx=0.1, Dy=2]{-}(0.00000, 0.00000)(0.00000, 0.00000)(0.30010, 10.00000)%
\multips(0.10000, 0.00000)(0.10000, 0.0){2}{\psline[linecolor=black, linestyle=dotted, linewidth=0.2mm](0, 0)(0, 10.00000)}
\multips(0.00000, 2.00000)(0, 2.00000){4}{\psline[linecolor=black, linestyle=dotted, linewidth=0.2mm](0, 0)(0.30010, 0)}
\rput[b](0.15005, -1.83673){$E_\mathrm{tx,norm}$}
\rput[t]{90}(-0.05456, 5.00000){$E_\mathrm{c,norm}$}
\psclip{\psframe(0.00000, 0.00000)(0.30010, 10.00000)}
\psline[linecolor=darkblue, plotstyle=curve, linewidth=0.2mm, showpoints=false, linestyle=solid, dotstyle=triangle, dotscale=1.7 1.7](0.24815, 0.35160)(0.10272, 0.52740)(0.05622, 0.70320)(0.03544, 0.87900)(0.02437, 1.05480)(0.01778, 1.23059)(0.01354, 1.40639)(0.01065, 1.58219)(0.00860, 1.75799)(0.00709, 1.93379)(0.00594, 2.10959)(0.00505, 2.28539)(0.00435, 2.46119)(0.00378, 2.63699)(0.00332, 2.81279)(0.00294, 2.98859)(0.00262, 3.16439)(0.00235, 3.34019)(0.00212, 3.51598)(0.00192, 3.69178)
\psline[linecolor=darkblue, plotstyle=curve, linewidth=0.2mm, showpoints=true, linestyle=none, dotstyle=triangle, dotscale=1.7 1.7](0.26563, 0.35160)(0.11312, 0.52740)(0.06297, 0.70320)(0.04015, 0.87900)(0.02784, 1.05480)(0.02044, 1.23059)(0.01564, 1.40639)(0.01235, 1.58219)(0.01000, 1.75799)(0.00827, 1.93379)(0.00695, 2.10959)(0.00592, 2.28539)(0.00510, 2.46119)(0.00445, 2.63699)(0.00391, 2.81279)(0.00346, 2.98859)(0.00309, 3.16439)(0.00277, 3.34019)(0.00250, 3.51598)(0.00227, 3.69178)
\psline[linecolor=darkgreen, plotstyle=curve, linewidth=0.2mm, showpoints=false, linestyle=solid, dotstyle=square, dotscale=1.7 1.7](0.23438, 2.00000)(0.10106, 3.00000)(0.05584, 4.00000)(0.03532, 5.00000)(0.02432, 6.00000)(0.01776, 7.00000)(0.01353, 8.00000)(0.01065, 9.00000)(0.00860, 10.00000)(0.00709, 11.00000)(0.00594, 12.00000)(0.00505, 13.00000)(0.00435, 14.00000)(0.00378, 15.00000)(0.00332, 16.00000)(0.00294, 17.00000)(0.00262, 18.00000)(0.00235, 19.00000)(0.00212, 20.00000)(0.00192, 21.00000)
\psline[linecolor=darkgreen, plotstyle=curve, linewidth=0.2mm, showpoints=true, linestyle=none, dotstyle=square, dotscale=1.7 1.7](0.25000, 2.00000)(0.11111, 3.00000)(0.06250, 4.00000)(0.04000, 5.00000)(0.02778, 6.00000)(0.02041, 7.00000)(0.01563, 8.00000)(0.01235, 9.00000)(0.01000, 10.00000)(0.00826, 11.00000)(0.00694, 12.00000)(0.00592, 13.00000)(0.00510, 14.00000)(0.00444, 15.00000)(0.00391, 16.00000)(0.00346, 17.00000)(0.00309, 18.00000)(0.00277, 19.00000)(0.00250, 20.00000)(0.00227, 21.00000)
\endpsclip
\psline[linecolor=black, linestyle=solid, linewidth=0.2mm]{->}(0.26500, 4.00000)(0.25000, 2.00000)
\psline[linecolor=black, linestyle=solid, linewidth=0.2mm]{->}(0.26500, 4.00000)(0.26563, 0.35160)
\psline[linecolor=black, linestyle=solid, linewidth=0.2mm]{->}(0.26563, 0.85160)(0.11312, 1.02740)
\rput[b](0.26500, 4.00000){\psframebox[linestyle=none, fillcolor=white, fillstyle=solid]{$N = 1$}}
\rput[c](0.17937, 1.60000){\psframebox[linestyle=none, fillcolor=white, fillstyle=solid]{\small{Increasing $N$}}}
\psframe[linecolor=black, fillstyle=solid, fillcolor=white, shadowcolor=lightgray, shadowsize=1mm, shadow=true](0.14732, 7.55102)(0.28373, 9.59184)
\rput[l](0.19643, 8.97959){$R_\mathrm{ref} = 2$}
\psline[linecolor=darkgreen, linestyle=solid, linewidth=0.3mm](0.15823, 8.97959)(0.18006, 8.97959)
\psline[linecolor=darkgreen, linestyle=solid, linewidth=0.3mm](0.15823, 8.97959)(0.18006, 8.97959)
\psdots[linecolor=darkgreen, linestyle=solid, linewidth=0.3mm, dotstyle=square, dotscale=1.7 1.7, linecolor=darkgreen](0.16915, 8.97959)
\rput[l](0.19643, 8.16327){$R_\mathrm{ref} = 0.1$}
\psline[linecolor=darkblue, linestyle=solid, linewidth=0.3mm](0.15823, 8.16327)(0.18006, 8.16327)
\psline[linecolor=darkblue, linestyle=solid, linewidth=0.3mm](0.15823, 8.16327)(0.18006, 8.16327)
\psdots[linecolor=darkblue, linestyle=solid, linewidth=0.3mm, dotstyle=triangle, dotscale=1.7 1.7, linecolor=darkblue](0.16915, 8.16327)
}\end{pspicture}
\endgroup
 
	\caption{Tradeoff of normalized computation and transmission energy for $N\in[1; 30]$ relays and
	reference rates $R_\mathrm{ref}$. Lines show the exact solution for wireless networks and markers show the exact solution for
	fixed networks (which also serves as lower bound for wireless networks). Furthermore, $\PathlossExponent=3$ and $\sigma^2=1$.
	}
	 \label{fig:tradeoff.region}
      \end{figure}
      Fig. \ref{fig:tradeoff.region} shows the achievable transmission-computation-energy curve for low-rate
      transmission ($R_\mathrm{ref} = 0.1$) and high-rate transmission ($R_\mathrm{ref} = 2$). Lines indicate the exact
      solution for wireless networks using $T_\mathrm{c, opt}'$ in (\ref{eq:protocol:390}) while markers show the solution
      for fixed networks derived in (\ref{eq:protocol:400}), which also serves as approximated solution for wireless networks.
      Each marker indicates one particular setup with $N$ relay nodes where higher $N$ result
      in lower normalized transmission-energy and higher normalized computation-energy. 
      
      The minimum normalized computation-energy is lower for the low-rate transmission than for the high-rate transmission, which
      indicates that in multi-hop networks with fixed $N$ the relative computation-energy savings are higher for low rates. 
      Consider the low-rate transmission and the slope of the curve. In the case of linear complexity, the computational energy
      doubles with every additional hop irrespective of the packet length. Hence, the normalized computational energy for a linearly
      complex algorithm has slope $1$ in $N$ and is significantly higher than for exponential complexity. On the other hand, for
      high-rate transmission the computational energy for exponential complexity also increases linearly in $N$ as $T_\mathrm{c, opt}'=T_\mathrm{ref}$.
      Therefore, low-rate transmission implies that coding with higher complexity is preferable over
      a less complex codes, while for high-rate transmission as in fixed networks less complex codes are preferable with respect
      to the transmission-computation-energy tradeoff.

    \subsection{Optimization of the Overall Energy}
      \begin{figure}
	\centering
	\subfigure[Network size]{\begingroup
\unitlength=1mm
\psset{xunit=10.71413mm, yunit=4.90000mm, linewidth=0.1mm}
\psset{arrowsize=2pt 3, arrowlength=1.4, arrowinset=.4}\psset{axesstyle=frame}
\begin{pspicture}(-1.12002, -2.24490)(7.00010, 10.00000)
\rput(-0.18667, -0.40816){%
\psaxes[subticks=0, labels=all, xsubticks=1, ysubticks=1, Ox=0, Oy=0, Dx=1, Dy=1]{-}(0.00000, 0.00000)(0.00000, 0.00000)(7.00010, 10.00000)%
\multips(1.00000, 0.00000)(1.00000, 0.0){6}{\psline[linecolor=black, linestyle=dotted, linewidth=0.2mm](0, 0)(0, 10.00000)}
\multips(0.00000, 1.00000)(0, 1.00000){9}{\psline[linecolor=black, linestyle=dotted, linewidth=0.2mm](0, 0)(7.00010, 0)}
\rput[b](3.50005, -1.83673){$R_\mathrm{ref}$}
\rput[t]{90}(-0.93335, 5.00000){Optimal $N$}
\psclip{\psframe(0.00000, 0.00000)(7.00010, 10.00000)}
\psline[linecolor=darkgreen, plotstyle=curve, linewidth=0.2mm, showpoints=false, linestyle=solid, dotstyle=square, dotscale=1.7 1.7](0.00010, 4.00000)(0.50010, 4.00000)(1.00010, 5.00000)(1.50010, 5.00000)(2.00010, 5.00000)(2.50010, 6.00000)(3.00010, 6.00000)(3.50010, 6.00000)(4.00010, 6.00000)(4.50010, 6.00000)(5.00010, 6.00000)(5.50010, 6.00000)(6.00010, 6.00000)(6.50010, 6.00000)(7.00010, 6.00000)
\psline[linecolor=darkgreen, plotstyle=curve, linewidth=0.2mm, showpoints=true, linestyle=none, dotstyle=square, dotscale=1.7 1.7](0.00010, 4.00000)(0.50010, 5.00000)(1.00010, 5.00000)(1.50010, 5.00000)(2.00010, 6.00000)(2.50010, 6.00000)(3.00010, 6.00000)(3.50010, 6.00000)(4.00010, 6.00000)(4.50010, 6.00000)(5.00010, 6.00000)(5.50010, 6.00000)(6.00010, 6.00000)(6.50010, 6.00000)(7.00010, 6.00000)
\psline[linecolor=darkgreen, plotstyle=curve, linewidth=0.2mm, showpoints=false, linestyle=dashed, dotstyle=square, dotscale=1.7 1.7](0.00010, 6.00000)(0.50010, 7.00000)(1.00010, 7.00000)(1.50010, 8.00000)(2.00010, 8.00000)(2.50010, 9.00000)(3.00010, 10.00000)(3.50010, 11.00000)(4.00010, 12.00000)(4.50010, 13.00000)(5.00010, 14.00000)(5.50010, 15.00000)(6.00010, 17.00000)(6.50010, 19.00000)(7.00010, 20.00000)
\psline[linecolor=darkgreen, plotstyle=curve, linewidth=0.2mm, showpoints=true, linestyle=none, dotstyle=square, dotscale=1.7 1.7](0.00010, 7.00000)(0.50010, 7.00000)(1.00010, 8.00000)(1.50010, 8.00000)(2.00010, 9.00000)(2.50010, 10.00000)(3.00010, 10.00000)(3.50010, 11.00000)(4.00010, 12.00000)(4.50010, 13.00000)(5.00010, 15.00000)(5.50010, 16.00000)(6.00010, 18.00000)(6.50010, 20.00000)(7.00010, 22.00000)
\psline[linecolor=darkred, plotstyle=curve, linewidth=0.2mm, showpoints=false, linestyle=solid, dotstyle=o, dotscale=1.7 1.7](0.00010, 2.00000)(0.50010, 2.00000)(1.00010, 2.00000)(1.50010, 2.00000)(2.00010, 2.00000)(2.50010, 2.00000)(3.00010, 2.00000)(3.50010, 2.00000)(4.00010, 2.00000)(4.50010, 2.00000)(5.00010, 2.00000)(5.50010, 2.00000)(6.00010, 2.00000)(6.50010, 2.00000)(7.00010, 2.00000)
\psline[linecolor=darkred, plotstyle=curve, linewidth=0.2mm, showpoints=true, linestyle=none, dotstyle=o, dotscale=1.7 1.7](0.00010, 2.00000)(0.50010, 2.00000)(1.00010, 2.00000)(1.50010, 2.00000)(2.00010, 2.00000)(2.50010, 2.00000)(3.00010, 2.00000)(3.50010, 2.00000)(4.00010, 2.00000)(4.50010, 2.00000)(5.00010, 2.00000)(5.50010, 2.00000)(6.00010, 2.00000)(6.50010, 2.00000)(7.00010, 2.00000)
\psline[linecolor=darkred, plotstyle=curve, linewidth=0.2mm, showpoints=false, linestyle=dashed, dotstyle=o, dotscale=1.7 1.7](0.00010, 2.00000)(0.50010, 3.00000)(1.00010, 3.00000)(1.50010, 3.00000)(2.00010, 3.00000)(2.50010, 4.00000)(3.00010, 4.00000)(3.50010, 5.00000)(4.00010, 5.00000)(4.50010, 5.00000)(5.00010, 6.00000)(5.50010, 7.00000)(6.00010, 7.00000)(6.50010, 8.00000)(7.00010, 9.00000)
\psline[linecolor=darkred, plotstyle=curve, linewidth=0.2mm, showpoints=true, linestyle=none, dotstyle=o, dotscale=1.7 1.7](0.00010, 3.00000)(0.50010, 3.00000)(1.00010, 3.00000)(1.50010, 3.00000)(2.00010, 4.00000)(2.50010, 4.00000)(3.00010, 4.00000)(3.50010, 5.00000)(4.00010, 5.00000)(4.50010, 6.00000)(5.00010, 6.00000)(5.50010, 7.00000)(6.00010, 8.00000)(6.50010, 9.00000)(7.00010, 9.00000)
\psline[linecolor=darkblue, plotstyle=curve, linewidth=0.2mm, showpoints=false, linestyle=solid, dotstyle=triangle, dotscale=1.7 1.7](0.00010, 1.00000)(0.50010, 1.00000)(1.00010, 1.00000)(1.50010, 0.00000)(2.00010, 1.00000)(2.50010, 1.00000)(3.00010, 1.00000)(3.50010, 1.00000)(4.00010, 1.00000)(4.50010, 1.00000)(5.00010, 1.00000)(5.50010, 1.00000)(6.00010, 1.00000)(6.50010, 1.00000)(7.00010, 1.00000)
\psline[linecolor=darkblue, plotstyle=curve, linewidth=0.2mm, showpoints=true, linestyle=none, dotstyle=triangle, dotscale=1.7 1.7](0.00010, 1.00000)(0.50010, 1.00000)(1.00010, 0.00000)(1.50010, 0.00000)(2.00010, 1.00000)(2.50010, 1.00000)(3.00010, 1.00000)(3.50010, 1.00000)(4.00010, 1.00000)(4.50010, 1.00000)(5.00010, 1.00000)(5.50010, 1.00000)(6.00010, 1.00000)(6.50010, 1.00000)(7.00010, 1.00000)
\psline[linecolor=darkblue, plotstyle=curve, linewidth=0.2mm, showpoints=false, linestyle=dashed, dotstyle=triangle, dotscale=1.7 1.7](0.00010, 1.00000)(0.50010, 1.00000)(1.00010, 1.00000)(1.50010, 1.00000)(2.00010, 1.00000)(2.50010, 1.00000)(3.00010, 1.00000)(3.50010, 2.00000)(4.00010, 2.00000)(4.50010, 2.00000)(5.00010, 2.00000)(5.50010, 3.00000)(6.00010, 3.00000)(6.50010, 3.00000)(7.00010, 4.00000)
\psline[linecolor=darkblue, plotstyle=curve, linewidth=0.2mm, showpoints=true, linestyle=none, dotstyle=triangle, dotscale=1.7 1.7](0.00010, 1.00000)(0.50010, 1.00000)(1.00010, 1.00000)(1.50010, 1.00000)(2.00010, 1.00000)(2.50010, 1.00000)(3.00010, 2.00000)(3.50010, 2.00000)(4.00010, 2.00000)(4.50010, 2.00000)(5.00010, 2.00000)(5.50010, 3.00000)(6.00010, 3.00000)(6.50010, 3.00000)(7.00010, 4.00000)
\endpsclip
}\end{pspicture}
\endgroup
 }
	\subfigure[Sum Energy normalized]{\begingroup
\unitlength=1mm
\psset{xunit=10.71413mm, yunit=24.50000mm, linewidth=0.1mm}
\psset{arrowsize=2pt 3, arrowlength=1.4, arrowinset=.4}\psset{axesstyle=frame}
\begin{pspicture}(-1.12002, -0.44898)(7.00010, 2.00000)
\rput(-0.18667, -0.08163){%
\psaxes[subticks=0, labels=all, xsubticks=1, ysubticks=1, Ox=0, Oy=0, Dx=1, Dy=0.5]{-}(0.00000, 0.00000)(0.00000, 0.00000)(7.00010, 2.00000)%
\multips(1.00000, 0.00000)(1.00000, 0.0){6}{\psline[linecolor=black, linestyle=dotted, linewidth=0.2mm](0, 0)(0, 2.00000)}
\multips(0.00000, 0.50000)(0, 0.50000){3}{\psline[linecolor=black, linestyle=dotted, linewidth=0.2mm](0, 0)(7.00010, 0)}
\rput[b](3.50005, -0.36735){$R_\mathrm{ref}$}
\rput[t]{90}(-0.93335, 1.00000){$E_\mathrm{sum, norm}$}
\psclip{\psframe(0.00000, 0.00000)(7.00010, 2.00000)}
\psline[linecolor=darkgreen, plotstyle=curve, linewidth=0.2mm, showpoints=false, linestyle=solid, dotstyle=square, dotscale=1.7 1.7](0.00010, 0.00141)(0.50010, 0.09073)(1.00010, 0.08007)(1.50010, 0.07018)(2.00010, 0.06389)(2.50010, 0.05991)(3.00010, 0.05743)(3.50010, 0.05584)(4.00010, 0.05480)(4.50010, 0.05409)(5.00010, 0.05361)(5.50010, 0.05328)(6.00010, 0.05304)(6.50010, 0.05288)(7.00010, 0.05277)
\psline[linecolor=darkgreen, plotstyle=curve, linewidth=0.2mm, showpoints=true, linestyle=none, dotstyle=square, dotscale=1.7 1.7](0.00010, 0.00148)(0.50010, 0.09445)(1.00010, 0.08349)(1.50010, 0.07361)(2.00010, 0.06663)(2.50010, 0.06254)(3.00010, 0.06006)(3.50010, 0.05848)(4.00010, 0.05744)(4.50010, 0.05673)(5.00010, 0.05625)(5.50010, 0.05591)(6.00010, 0.05568)(6.50010, 0.05552)(7.00010, 0.05540)
\psline[linecolor=darkgreen, plotstyle=curve, linewidth=0.2mm, showpoints=false, linestyle=dashed, dotstyle=square, dotscale=1.7 1.7](0.00010, 0.15490)(0.50010, 0.13768)(1.00010, 0.12156)(1.50010, 0.10663)(2.00010, 0.09305)(2.50010, 0.08043)(3.00010, 0.06925)(3.50010, 0.05930)(4.00010, 0.05050)(4.50010, 0.04279)(5.00010, 0.03609)(5.50010, 0.03034)(6.00010, 0.02539)(6.50010, 0.02120)(7.00010, 0.01763)
\psline[linecolor=darkgreen, plotstyle=curve, linewidth=0.2mm, showpoints=true, linestyle=none, dotstyle=square, dotscale=1.7 1.7](0.00010, 0.16248)(0.50010, 0.14423)(1.00010, 0.12776)(1.50010, 0.11195)(2.00010, 0.09764)(2.50010, 0.08473)(3.00010, 0.07296)(3.50010, 0.06247)(4.00010, 0.05323)(4.50010, 0.04517)(5.00010, 0.03810)(5.50010, 0.03203)(6.00010, 0.02683)(6.50010, 0.02240)(7.00010, 0.01864)
\psline[linecolor=darkred, plotstyle=curve, linewidth=0.2mm, showpoints=false, linestyle=solid, dotstyle=o, dotscale=1.7 1.7](0.00010, 0.00071)(0.50010, 0.37888)(1.00010, 0.34948)(1.50010, 0.30918)(2.00010, 0.28224)(2.50010, 0.26705)(3.00010, 0.25776)(3.50010, 0.25179)(4.00010, 0.24784)(4.50010, 0.24517)(5.00010, 0.24334)(5.50010, 0.24207)(6.00010, 0.24119)(6.50010, 0.24057)(7.00010, 0.24014)
\psline[linecolor=darkred, plotstyle=curve, linewidth=0.2mm, showpoints=true, linestyle=none, dotstyle=o, dotscale=1.7 1.7](0.00010, 0.00072)(0.50010, 0.38770)(1.00010, 0.35876)(1.50010, 0.31851)(2.00010, 0.29166)(2.50010, 0.27652)(3.00010, 0.26727)(3.50010, 0.26132)(4.00010, 0.25738)(4.50010, 0.25472)(5.00010, 0.25290)(5.50010, 0.25163)(6.00010, 0.25076)(6.50010, 0.25014)(7.00010, 0.24971)
\psline[linecolor=darkred, plotstyle=curve, linewidth=0.2mm, showpoints=false, linestyle=dashed, dotstyle=o, dotscale=1.7 1.7](0.00010, 0.65751)(0.50010, 0.58839)(1.00010, 0.52416)(1.50010, 0.46646)(2.00010, 0.41554)(2.50010, 0.36073)(3.00010, 0.31257)(3.50010, 0.27131)(4.00010, 0.23075)(4.50010, 0.19764)(5.00010, 0.16637)(5.50010, 0.14045)(6.00010, 0.11784)(6.50010, 0.09831)(7.00010, 0.08185)
\psline[linecolor=darkred, plotstyle=curve, linewidth=0.2mm, showpoints=true, linestyle=none, dotstyle=o, dotscale=1.7 1.7](0.00010, 0.67703)(0.50010, 0.60718)(1.00010, 0.54330)(1.50010, 0.48592)(2.00010, 0.43107)(2.50010, 0.37478)(3.00010, 0.32676)(3.50010, 0.28188)(4.00010, 0.24140)(4.50010, 0.20589)(5.00010, 0.17462)(5.50010, 0.14700)(6.00010, 0.12357)(6.50010, 0.10349)(7.00010, 0.08626)
\psline[linecolor=darkblue, plotstyle=curve, linewidth=0.2mm, showpoints=false, linestyle=solid, dotstyle=triangle, dotscale=1.7 1.7](0.00010, 0.00041)(0.50010, 0.93084)(1.00010, 1.06226)(1.50010, 1.00000)(2.00010, 0.94062)(2.50010, 0.90154)(3.00010, 0.87642)(3.50010, 0.85977)(4.00010, 0.84850)(4.50010, 0.84078)(5.00010, 0.83544)(5.50010, 0.83172)(6.00010, 0.82911)(6.50010, 0.82728)(7.00010, 0.82600)
\psline[linecolor=darkblue, plotstyle=curve, linewidth=0.2mm, showpoints=true, linestyle=none, dotstyle=triangle, dotscale=1.7 1.7](0.00010, 0.00041)(0.50010, 0.93715)(1.00010, 1.06655)(1.50010, 1.00000)(2.00010, 0.94999)(2.50010, 0.91126)(3.00010, 0.88636)(3.50010, 0.86986)(4.00010, 0.85869)(4.50010, 0.85104)(5.00010, 0.84574)(5.50010, 0.84205)(6.00010, 0.83947)(6.50010, 0.83766)(7.00010, 0.83639)
\psline[linecolor=darkblue, plotstyle=curve, linewidth=0.2mm, showpoints=false, linestyle=dashed, dotstyle=triangle, dotscale=1.7 1.7](0.00010, 1.48464)(0.50010, 1.42963)(1.00010, 1.37057)(1.50010, 1.30846)(2.00010, 1.24468)(2.50010, 1.18088)(3.00010, 1.11880)(3.50010, 0.99869)(4.00010, 0.88385)(4.50010, 0.78121)(5.00010, 0.69184)(5.50010, 0.59921)(6.00010, 0.50904)(6.50010, 0.43542)(7.00010, 0.36704)
\psline[linecolor=darkblue, plotstyle=curve, linewidth=0.2mm, showpoints=true, linestyle=none, dotstyle=triangle, dotscale=1.7 1.7](0.00010, 1.50487)(0.50010, 1.45202)(1.00010, 1.39527)(1.50010, 1.33560)(2.00010, 1.27433)(2.50010, 1.21303)(3.00010, 1.14593)(3.50010, 1.02243)(4.00010, 0.90895)(4.50010, 0.80753)(5.00010, 0.71921)(5.50010, 0.61794)(6.00010, 0.52827)(6.50010, 0.45505)(7.00010, 0.38106)
\endpsclip
\psframe[linecolor=black, fillstyle=solid, fillcolor=white, shadowcolor=lightgray, shadowsize=1mm, shadow=true](3.08004, 1.46939)(6.53343, 2.04082)
\rput[l](3.92006, 1.91837){$\eta_\mathrm{ref}(1) = \unit[0]{dB}$}
\psline[linecolor=darkblue, linestyle=solid, linewidth=0.3mm](3.26671, 1.91837)(3.64005, 1.91837)
\psline[linecolor=darkblue, linestyle=solid, linewidth=0.3mm](3.26671, 1.91837)(3.64005, 1.91837)
\psdots[linecolor=darkblue, linestyle=solid, linewidth=0.3mm, dotstyle=triangle, dotscale=1.7 1.7, linecolor=darkblue](3.45338, 1.91837)
\rput[l](3.92006, 1.75510){$\eta_\mathrm{ref}(1) = \unit[-10]{dB}$}
\psline[linecolor=darkred, linestyle=solid, linewidth=0.3mm](3.26671, 1.75510)(3.64005, 1.75510)
\psline[linecolor=darkred, linestyle=solid, linewidth=0.3mm](3.26671, 1.75510)(3.64005, 1.75510)
\psdots[linecolor=darkred, linestyle=solid, linewidth=0.3mm, dotstyle=o, dotscale=1.7 1.7, linecolor=darkred](3.45338, 1.75510)
\rput[l](3.92006, 1.59184){$\eta_\mathrm{ref}(1) = \unit[-20]{dB}$}
\psline[linecolor=darkgreen, linestyle=solid, linewidth=0.3mm](3.26671, 1.59184)(3.64005, 1.59184)
\psline[linecolor=darkgreen, linestyle=solid, linewidth=0.3mm](3.26671, 1.59184)(3.64005, 1.59184)
\psdots[linecolor=darkgreen, linestyle=solid, linewidth=0.3mm, dotstyle=square, dotscale=1.7 1.7, linecolor=darkgreen](3.45338, 1.59184)
}\end{pspicture}
\endgroup
 }
	\caption{Minimum normalized sum-energy and the optimal number of relay nodes depending on the reference rate $R_\mathrm{ref}$.
	Markers show the solution for fixed networks and lines show the solution for wireless networks. Dashed lines
	give the results for linear complexity while solid lines give the results for exponential complexity.
	Again, $\PathlossExponent=3$, $\sigma^2=1$.}
	\label{fig:results:optimal.network.parameters}
      \end{figure}
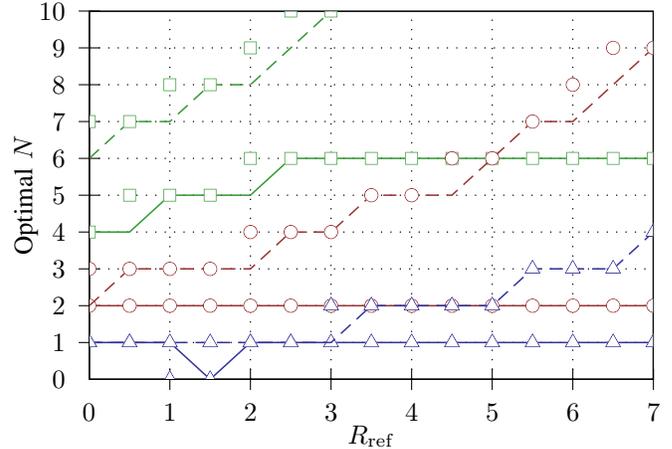
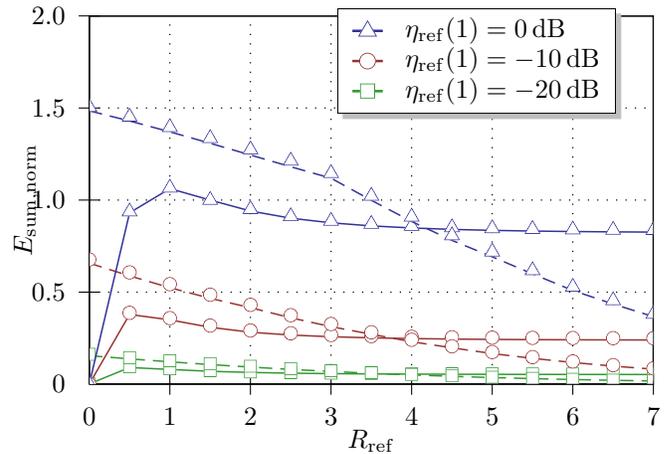
      Fig. \ref{fig:results:optimal.network.parameters} shows
      the normalized sum-energy according to (\ref{eq:protocol:590}) for three different values of $\eta_\mathrm{ref}(1)$ and the optimal number of relay nodes, which
      minimizes the sum-energy for a particular value of $\eta_\mathrm{ref}(1)$ and $R_\mathrm{ref}$ and $T_{c,opt}$ given in (\ref{eq:protocol:520}).
      In Fig. \ref{fig:results:optimal.network.parameters}, solid lines indicate the results for exponential complexity
      and dashed lines for linear complexity relative to the same common reference power 
      (after application of a $\unit[5]{dB}$ SNR gap). Markers again show the solution for fixed networks while lines show the solution for wireless networks.  

      Fig. \ref{fig:results:optimal.network.parameters}(a) shows the optimal $N$ depending on $\eta_\mathrm{ref}(1)$ and
      $R_\mathrm{ref}$. The higher the computation energy compared to the transmission energy (reflected by a higher $\eta_\mathrm{ref}(1)$)
      the lower the optimal $N$. With an increasing emphasis on the computation energy, it becomes the dominant part of
      the sum-energy, which renders a higher number of nodes less beneficial. 
      In addition, if the reference rate is increasing, the optimal number of nodes is also increasing in order to counteract the exponentially increasing
      transmission energy. The slope of this increase is higher for $\eta_\mathrm{ref}(1) = \unit[-20]{dB}$ than for $\eta_\mathrm{ref}(1) = \unit[0]{dB}$.
      The latter refers to fixed networks where computation energy contributes more significantly to the sum-energy than in wireless networks. In addition,
      networks with linear computational complexity prefer more hops than networks with exponential complexity.
      Further consider the optimal $N$ at $R_\mathrm{ref} = 1.5$ for $\eta_\mathrm{ref}(1) = \unit[0]{dB}$. It reaches an minimum at this point as the normalized
      sum-energy is greater than $1$ and therefore relaying is not optimal for this case. However, this is different for linear complexity as we apply
      a $\unit[5]{dB}$ shift, which then renders relaying beneficial again.

      Fig. \ref{fig:results:optimal.network.parameters}(b) shows the minimum normalized sum-energy using 
      $T_{c, opt}'$ in (\ref{eq:protocol:520}) and using the optimal $N$ depicted 
      in Fig. \ref{fig:results:optimal.network.parameters}(a). The lowest value of $E_\mathrm{sum, norm}$ is obtained
      for $\eta_\mathrm{ref}(1) = \unit[-20]{dB}$ as the transmission energy can be significantly reduced and the computation energy does not
      become a dominating part with increasing $N$. With increasing reference rates, the transmission energy in the single-hop network 
      becomes a more dominant part of the sum-energy. Due to the significant transmission power savings in multi-hop networks, also 
      the normalized sum-energy declines with increasing $R_\mathrm{ref}$.
      This implies that multi-hop transmission
      is more useful in scenarios with high data rates and less complex decoders and encoders. 
      We can further see that at low rates the sum-energy is higher for linear complexity than for exponential complexity while at higher rates
      linear complexity is again preferable with respect to the sum-energy. By contrast,
      for $\eta_\mathrm{ref}(1) = \unit[-20]{dB}$ (wireless case)
      both linear and exponential complexity achieve similar sum-energy performance (as the transmission power is dominating).

  \section{Conclusions and Future Challenges}\label{sec:conclusions}
    This paper introduced and analyzed the tradeoff of the energy required for decoding and processing transmissions and the
    energy necessary to transmit a message. We derived a framework, which showed that for increasing emphasis on the computation energy 
    (increasing $\eta_\mathrm{ref}(1)$), multi-hop protocols are less beneficial to reduce the network-wide spent
    energy, while for increasing emphasis on the transmission energy (increasing reference rate $R_\mathrm{ref}$) they become
    more beneficial. The comparison of linear and exponential complexity showed that more complex encoding is preferable at 
    low data rates while low complex encoding is preferable at high rates with respect to the transmission-computation-energy tradeoff.
    In addition, using different weighting of the transmission and computation energy as for instance in wireless and fixed networks,
    we showed that a smaller number of hops in fixed networks is preferable due to the significant computation energy while
    in wireless networks more hops are preferable due to the dominating transmission energy. 
    
    Among the next challenges is the question for the optimal complexity-function rather than for the optimal protocol or number of hops. 
    If a functional expression of the SNR-gap depending on the computational complexity can be found, 
    the optimal computation complexity for both wireless and fixed networks can be determined.
    
  \appendix
  In section \ref{sec:protocol:optimum}, we introduced the function $\eta_\mathrm{ref}(R_\mathrm{ref})$, which relates the
  computation and transmission energy in the single-hop reference system 
  such that $E_\mathrm{c, ref} = \eta_\mathrm{ref}(R_\mathrm{ref})E_\mathrm{tx, ref}$.
  Assume that for reference rate $R_\mathrm{ref} = 1$
  the function is predefined as a system-specific parameter. At this rate the source must transmit on the direct link with
  power $P_\mathrm{tx, ref}(1)$. Assume that the rate is now given by $R' = \delta_r R_\mathrm{ref}$, then the transmission energy is
  given by $E_\mathrm{tx, ref}(1/\delta_r)$, where we applied (\ref{eq:model:energy:tx.10}). 
  The computation energy is given by $2^{\Delta_r}E_\mathrm{c, ref}(1)$ as given by (\ref{eq:model:energy:tx.11}).
  Hence, we can derive $\eta_\mathrm{ref}(R_\mathrm{ref})$ as
  \begin{align}
    2^{\Delta_r}E_\mathrm{c, ref}(1) & = \eta_\mathrm{ref}(R_\mathrm{ref})\cdot E_\mathrm{tx, ref}(1/\delta_r) \\
    2^{\Delta_r}\eta_\mathrm{ref}(1)E_\mathrm{tx, ref}(1) & = \eta_\mathrm{ref}(R_\mathrm{ref})\cdot E_\mathrm{tx, ref}(1/\delta_r) \\
    \eta_\mathrm{ref}(R_\mathrm{ref}) & =
      \eta_\mathrm{ref}(1)\cdot\frac{2^{\Delta_r}E_\mathrm{tx, ref}(1)}{E_\mathrm{tx, ref}(1/\delta_r)}.
  \end{align}

  \bibliographystyle{IEEEtran}
  \bibliography{IEEEfull,my-references}

\begin{thebibliography}{1}
\providecommand{\url}[1]{#1}
\csname url@rmstyle\endcsname
\providecommand{\newblock}{\relax}
\providecommand{\bibinfo}[2]{#2}
\providecommand\BIBentrySTDinterwordspacing{\spaceskip=0pt\relax}
\providecommand\BIBentryALTinterwordstretchfactor{4}
\providecommand\BIBentryALTinterwordspacing{\spaceskip=\fontdimen2\font plus
\BIBentryALTinterwordstretchfactor\fontdimen3\font minus
  \fontdimen4\font\relax}
\providecommand\BIBforeignlanguage[2]{{%
\expandafter\ifx\csname l@#1\endcsname\relax
\typeout{** WARNING: IEEEtran.bst: No hyphenation pattern has been}%
\typeout{** loaded for the language `#1'. Using the pattern for}%
\typeout{** the default language instead.}%
\else
\language=\csname l@#1\endcsname
\fi
#2}}

\bibitem{Avestimehr.Tse.TransIT.2007}
A.~Avestimehr and D.~Tse, ``Outage capacity of the fading relay channel in the
  low-{SNR} regime,'' \emph{{IEEE} Transactions on Information Theory},
  vol.~53, no.~4, pp. 1401--1415, April 2007.

\bibitem{Bhardwaj.Chandrakasan.ICC.2001}
M.~Bhardwaj and A.~Chandrakasan, ``Upper bounds on the lifetime of sensor
  networks,'' in \emph{{IEEE} International Conference on Communications},
  Helsinki, Finland, June 2001.

\bibitem{Melo.Liu.Globecom.2002}
E.~Duarte-Melo and M.~Liu, ``Analysis of energy consumption and lifetime of
  heterogeneous wireless sensor networks,'' in \emph{{IEEE} Global
  Communications Conference}, Taipei, Taiwan, November 2002.

\bibitem{Heinzelman.Chandrakasan.Balakrishnan.HICSS.2000}
W.~Heinzelman, A.~Chandrakasan, and H.~Balakrishnan, ``Energy-efficient
  communication protocol for wireless microsensor networks,'' in \emph{33rd
  Hawaii International Conference on System Sciences}, Hawaii, USA, January
  2000.

\bibitem{Tarokh.Seshadri.Calderbank.TransIT.1998}
V.~Tarokh, N.~Seshadri, and A.~Calderbank, ``Space-time codes for high data
  rate wireless communications: Performance criterion and code construction,''
  \emph{{IEEE} Transactions on Information Theory}, vol.~44, no.~2, pp.
  744--765, March 1998.

\bibitem{Li.Stuber.2006}
B.~Li and G.~St\"uber, \emph{Orthogonal Frequency Division Multiplexing for
  Wireless Communications}.\hskip 1em plus 0.5em minus 0.4em\relax
  Birkh\"auser, 2006.

\bibitem{Yu.Varodayan.Cioffi.TransIT.2005}
W.~Yu, D.~Varodayan, and J.~Cioffi, ``Trellis and convolutional precoding for
  transmitter-based interference presubtraction,'' \emph{{IEEE} Transactions on
  Information Theory}, vol.~53, no.~7, pp. 1220--1230, July 2005.

\bibitem{Xie.Kumar.TransIT.2004}
L.-L. Xie and P.~Kumar, ``A network information theory for wireless
  communication: Scaling laws and optimal operation,'' \emph{{IEEE}
  Transactions on Information Theory}, vol.~50, no.~5, pp. 748--767, May 2004.

\end{thebibliography}
\end{document}